\pdfoutput=1

\documentclass{article}
\usepackage{arxiv}
\usepackage[english]{babel}
\usepackage{nicefrac}    
\usepackage{microtype}   
\usepackage{booktabs}    
\usepackage[numbers]{natbib}
\usepackage{doi}
\usepackage{adjustbox}
\usepackage{float}
\usepackage[T1]{fontenc}
\usepackage{graphicx}
\usepackage{hhline}
\usepackage[utf8]{inputenc}
\usepackage{multicol}
\usepackage{multirow}

\usepackage{tikz}
\def\checkmark{\tikz\fill[scale=0.4](0,.35) -- (.25,0) -- (1,.7) -- (.25,.15) -- cycle;}

\title{Exploring Student Engagement and Outcomes: Experiences from Three Cycles of an Undergraduate Module}

\author{ \href{https://orcid.org/0000-0001-5066-425X}{\includegraphics[scale=0.06]{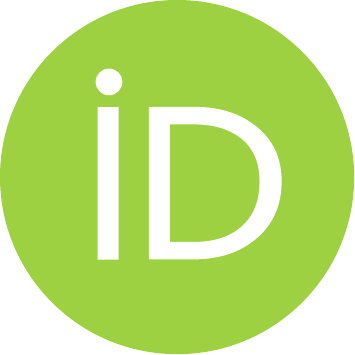}\hspace{1mm}Robert D.~Macredie} \\
	Department of Computer Science\\
	Brunel University London\\
	London, United Kingdom \\
	\texttt{rob.macredie@brunel.ac.uk} \\
	\And
	\href{https://orcid.org/0000-0003-1874-6145}{\includegraphics[scale=0.06]{orcid.pdf}\hspace{1mm}Martin Shepperd} \\
	Department of Computer Science\\
	Brunel University London\\
	London, United Kingdom \\
	\texttt{martin.shepperd@brunel.ac.uk} \\
	\And
	\href{https://orcid.org/0000-0001-6826-9688}{\includegraphics[scale=0.06]{orcid.pdf}\hspace{1mm}Tommaso Turchi} \\
	Department of Computer Science\\
	University of Pisa\\
	Pisa, Italy \\
	\texttt{tommaso.turchi@unipi.it} \\
	\And
	\href{https://orcid.org/0000-0001-9545-2269}{\includegraphics[scale=0.06]{orcid.pdf}\hspace{1mm}Terry Young} \\
	Brunel University London\\
	London, United Kingdom \\
	\texttt{terence.p.young@gmail.com} \\
}

\renewcommand{\headeright}{}
\renewcommand{\undertitle}{}
\renewcommand{\shorttitle}{Exploring Student Engagement and Outcomes}

\hypersetup{
pdftitle={Exploring Student Engagement and Outcomes: Experiences from Three Cycles of an Undergraduate Module},
pdfsubject={q-bio.NC, q-bio.QM}, 
pdfauthor={},
pdfkeywords={},
}

\begin{document}
\maketitle

\begin{abstract}
Many studies in educational data mining address specific learner groups, such as first-in-family to attend Higher Education, or focus on differences in characteristics such as gender or ethnicity, with the aim of predicting performance and designing interventions to improve outcomes. For Higher Education, this is reflected in significant interest in institutional-level analysis of student cohorts and in tools being promoted to Higher Education Institutions to support collection, integration and analysis of data. For those leading modules/units on degree programmes, however, the reality can be far removed from the seemingly well-supported and increasingly sophisticated approaches advocated in centrally-led data analysis. Module leaders often find themselves working with a number of student-data systems that are not integrated, may contain conflicting data and where significant effort is required to extract, clean and meaningfully analyse the data. This paper suggests that important lessons may be learned from experiences at module level in this context and from subsequent analysis of related data extracted/collected across multiple years. To explore this contention, the paper focuses on: (i) whether the likelihood of students failing can be predicted through engagement data; (ii) whether engagement is linked to final achievement and, if so, whether final grade may be predicted from the way students approached the module; (iii) how failing the first submission of a piece of coursework impacts the final outcome; and (iv) whether any identified patterns in the data are stable across instances of the module. These issues are analysed through data gathered from the module over a three-year period, where changes were made each year. The changes are described and a range of data analysis methods are applied, post hoc, to identify findings in relation to the four areas of focus. The key findings are that non-engagement with the Virtual Learning Environment in the first three weeks was the strongest predictor of failure and that early engagement correlated most strongly with final grade. General recommendations are drawn from the findings which should be valuable to module leaders in environments where access to integrated, up-to-date student information remains a day-to-day challenge, and insights will be presented into how such bottom-up activities might inform institutional/top-down planning in the use of relevant technologies.
\end{abstract}


\section{Introduction}

Most campuses have a Virtual Learning Environment (VLE) or Learning Management System (LMS), from which data is mined to make early predictions of student failure \cite{macfadyen2010mining} \cite{hu2014developing} or success \cite{pistilli2010purdue}, or to identify those ‘at risk’ of leaving a programme \cite{defreitas2015foundations} \cite{marbouti2016models}. 

   Effective predictions require data gathered from sources beyond the VLE/LMS, such as in-class clickers \cite{choi2018learning}, including coursework or test outcomes, all integrated into a reliable picture of an individual student’s progress \cite{mangaroska2017learning} in order to realise the potential benefits in prediction, retention, and outcomes \cite{yau2018using}. There is a burgeoning literature on the use of educational data initiatives \cite{na2017a}, including initiatives at Purdue and the Open University \cite{arnold2012course} \cite{kuzilek2015ou}, as well as on-line assessment \cite{brady2019academic}, alongside interest in assessing different elements of delivery in the light of student performance \cite{holmes2019learning}. Underpinning this is a literature on algorithms and the interventions based on them \cite{sonderlund2018the}, in turn part of a wider agenda around educational technology and transformation, of which blended learning and Massive Open Online Courses (MOOCs) are prominent examples.

     Recent evidence raises issues about the pace, at least, of transformation. \cite{gaebel2018learning}, for instance, note tensions between top-down and bottom-up design processes, while others (such as \cite{walker2012the}) identify a gap between the rhetoric of technology-supported pedagogy and the reality of the systems being installed, with institutions investing for efficiency gains and better student survey results at a time when there is limited innovation and serious evaluation at the level of academic programme units or modules. There are concerns that technology usage, which is easier to measure, is being confused with, or used as a proxy for engagement, which is predictive of grades \cite{dunn2019technology}. Finally, there are complaints that technology is not well aligned with pedagogic theory \cite{vieira2018visual}, with \cite{jansen2019a}noting that the technological agenda tends to dominate the pedagogical, while  \cite{jivet2017awareness} report that ``current designs foster competition between learners rather than knowledge mastery". 

   Delivering improved efficiency and student outcomes while addressing these issues is therefore a challenge for individuals and local teams in academic units, who must find a way to exploit the potential of VLEs and LMSs both to innovate and to evaluate innovation. 

This paper reports experiences at module level of using a VLE/LMS together with institutional and ‘homemade’ systems, to identify lessons learned in a \textit{post hoc} analysis of data collected over several years. It aims to provide guidance on engagement, retention, and outcomes, to others running modules. Further, it offers insight into how such bottom-up activity and analysis can inform top-down planning and institutional adoption of such systems.

To this end, we describe the introduction of blended learning to a final-year, undergraduate (Bachelors) module. The data were initially collected on an on-going basis from several sources, including student attendance at lectures and seminars and engagement with the VLE, and were used to identify and respond to needs and to make changes to the module from year to year. At the end of three years, there was an opportunity for retrospective, in-depth analysis in order to propose specific early warning metrics and to explore how different elements of the delivery (on-line, early engagement, late engagement) correlated with the resulting student outcomes.

This paper sets this study of blended learning in a theoretical context and presents a framework within which innovators at the module level may develop their pedagogy and use appropriate measures to assess their progress. From there, it develops a set of research questions that relate, firstly, to the wider applicability of this type of research and then to the specifics of quantifiable findings. Next, is a description of the module development, the data collected and the way in which datasets over three years were analysed, together with the findings. Finally, the discussion returns to the research questions to discuss the extent to which we believe the approach could be replicated elsewhere and how generalisable the findings may be. 

\section{Methods and Theory}

Although this initiative took place within a specific context, the COVID-19 pandemic has created the need for many groups to adopt online or blended learning, within even shorter timescales, needing to preserve existing standards while delivering learning and teaching using different platforms, and then to evaluate their success in doing so as a basis from which to refine subsequent activities.

In the case reported in this paper, and under departmental governance, introducing new technology was possible so long as the learning outcomes, core content and assessment approaches remained unchanged while variations in the amount of student-facing time were strictly managed. Experiments based on splitting the class or double teaching using diverse methods to assess changes in a more controlled way, which would have allowed clearer comparisons between cohorts within a year, were therefore not option.

Though not unique (see, for example the study by \cite{boulton2018virtual}), the \textit{post hoc} approach reported in this paper contrasts with many studies in this area of which we are aware, where evidence has been gained through literature reviews \cite{yau2018using} \cite{bodily2018open} \cite{papamitsiou2014learning}, independent studies around existing teaching and learning \cite{stricker2011efficient} \cite{wang2017an}, or trials, often with relatively small numbers of students \cite{raes2020learning} \cite{vanraaij2008the} \cite{mohamed2018implementing} \cite{chis2018investigating}. Some research has used VLE presence/absence as a surrogate for engagement or self-reported measures of engagement or achievement \cite{murillozamorano2019how}, which may or may not reflect a more objective assessment \cite{boulton2018virtual} \cite{raes2020learning}. There have also been studies that report both \cite{segers2001new}. 

       Given the constraints in relation to continuity from year to year, we adopted a form of Design Research \cite{collins2004design}, by having a clear goal in view and using what data could be gathered in approaching that goal. We applied a constructivist model of self-regulated learning \cite{winne2006how}. In brief, constructivism is a learning theory stating that learners construct new knowledge through experiences and reflection. Thus, this kind of pedagogy strives to provide students with experiential learning techniques through real-world problem solving and enables them to construct their knowledge through discussion, collaboration, and reflection. Indeed, Problem-Based Learning is regarded by some \cite{schmidt2009constructivist} as an example of a cognitive constructivist approach to education, whose goal is to help students build mental models of the world.

  The teaching and learning package resulting from our study is similar to that reported by \cite{jovanovic2019predictive}, whose focus is on first, rather than final year, engineers, rather than computer scientists. As with ours, their study covers three years, although the total cohort of students is more than twice as large.

 The final iteration of the teaching and learning package was a mix of classic blended learning and problem-based learning \cite{ayala2019understanding}, the latter often not being included in purist blended learning \cite{mojtahedi2019case}. The connection between the two was the assessment scenario – an open ended task to price up a software project – which provided a scaffold for learning in the form of a structured (but open-ended) set of documents for assessment, constructed by each student, from planning a project to delivering a final quotation to undertake the project work – following \cite{kelley2019the}. We also used light-touch tests \cite{chis2018investigating} before and after on-line study as a way to establish progress and to augment other tracked measures (such as attendance in person at seminars and lectures) (for more detail see section 3). 

  The pedagogy itself was a mix of classic blended learning – personal study followed by group discussion – with problem-based learning, thus avoiding the issue of lack of application. To realise this end, we designed a phase of mainstream problem-based learning \cite{ayala2019understanding} \cite{vandervleuten2019assessment} \cite{zwaal2019assessment} that consolidated the learning and provided a means of assessment. Table 1 shows the main elements of learning and teaching delivery as reported in recent research, compared with the implementation reported in this study to position our work. 

\begin{table}[!htbp]
\renewcommand{\arraystretch}{1.3}
\begin{adjustbox}{max width=\textwidth}
\begin{tabular}{p{4.65cm}p{1.68cm}p{1.73cm}p{1.73cm}p{1.72cm}p{4.74cm}}
\hhline{~----~}
\multicolumn{1}{p{4.75cm}}{} & 
\multicolumn{4}{|p{6.67cm}}{Recent Research Studies} & 
\multicolumn{1}{|p{4.82cm}}{} \\ 
\hline
\multicolumn{1}{|p{4.75cm}}{Literature Theme} & 
\multicolumn{1}{|p{1.5cm}}{\cite{vandervleuten2019assessment}} & 
\multicolumn{1}{|p{1.73cm}}{\cite{zwaal2019assessment}} & 
\multicolumn{1}{|p{1.73cm}}{\cite{mojtahedi2019case}} & 
\multicolumn{1}{|p{1.73cm}}{\cite{jovanovic2019predictive}} & 
\multicolumn{1}{|p{4.82cm}|}{Implementation in This Study} \\ 
\hline
\multicolumn{1}{|p{4.75cm}}{Private study for rote learning} & 
\multicolumn{1}{|p{1.5cm}}{} & 
\multicolumn{1}{|p{1.73cm}}{} & 
\multicolumn{1}{|p{1.73cm}}{\checkmark} & 
\multicolumn{1}{|p{1.73cm}}{\checkmark} & 
\multicolumn{1}{|p{4.82cm}|}{On-line 'mini-modules’ consisting of reading, watching lectures and simple quizzes to track engagement. } \\ 
\hline
\multicolumn{1}{|p{4.75cm}}{Face-to-face for active learning} & 
\multicolumn{1}{|p{1.5cm}}{} & 
\multicolumn{1}{|p{1.73cm}}{} & 
\multicolumn{1}{|p{1.73cm}}{\checkmark} & 
\multicolumn{1}{|p{1.73cm}}{\checkmark} & 
\multicolumn{1}{|p{4.82cm}|}{} \\ 
\hline
\multicolumn{1}{|p{4.75cm}}{On-Line quizzes} & 
\multicolumn{1}{|p{1.5cm}}{} & 
\multicolumn{1}{|p{1.73cm}}{} & 
\multicolumn{1}{|p{1.73cm}}{\checkmark} & 
\multicolumn{1}{|p{1.73cm}}{\checkmark} & 
\multicolumn{1}{|p{4.82cm}|}{On-line, rather than in class. } \\ 
\hline
\multicolumn{1}{|p{4.75cm}}{Peer group assessment} & 
\multicolumn{1}{|p{1.5cm}}{} & 
\multicolumn{1}{|p{1.73cm}}{} & 
\multicolumn{1}{|p{1.73cm}}{\checkmark} & 
\multicolumn{1}{|p{1.73cm}}{} & 
\multicolumn{1}{|p{4.82cm}|}{Formative only, all-to-all. } \\ 
\hline
\multicolumn{1}{|p{4.75cm}}{Tutor as facilitator or guide or assessor} & 
\multicolumn{1}{|p{1.5cm}}{\checkmark} & 
\multicolumn{1}{|p{1.73cm}}{\checkmark} & 
\multicolumn{1}{|p{1.73cm}}{\checkmark} & 
\multicolumn{1}{|p{1.73cm}}{\checkmark} & 
\multicolumn{1}{|p{4.82cm}|}{Tutors played several roles: curators of on-line material, facilitators of discussions, and assessors (formative and summative).} \\ 
\hline
\multicolumn{1}{|p{4.75cm}}{Student-centred learning: self-directed; self-regulated} & 
\multicolumn{1}{|p{1.5cm}}{\checkmark} & 
\multicolumn{1}{|p{1.73cm}}{\checkmark} & 
\multicolumn{1}{|p{1.73cm}}{} & 
\multicolumn{1}{|p{1.73cm}}{\checkmark} & 
\multicolumn{1}{|p{4.82cm}|}{On-line and through addressing the assessment scenario. } \\ 
\hline
\multicolumn{1}{|p{4.75cm}}{Group working} & 
\multicolumn{1}{|p{1.5cm}}{\checkmark} & 
\multicolumn{1}{|p{1.73cm}}{\checkmark} & 
\multicolumn{1}{|p{1.73cm}}{} & 
\multicolumn{1}{|p{1.73cm}}{} & 
\multicolumn{1}{|p{4.82cm}|}{Groups encouraged, but submissions had to be an individual's work.} \\ 
\hline
\multicolumn{1}{|p{4.75cm}}{Competence-based learning from real-world examples} & 
\multicolumn{1}{|p{1.5cm}}{\checkmark} & 
\multicolumn{1}{|p{1.73cm}}{\checkmark} & 
\multicolumn{1}{|p{1.73cm}}{} & 
\multicolumn{1}{|p{1.73cm}}{} & 
\multicolumn{1}{|p{4.82cm}|}{Scenario typically build with an industry expert.} \\ 
\hline
\end{tabular}
\end{adjustbox}
\caption{Main elements of learning and teaching delivery}
\label{tab:main_elements_learning_and_teaching}\end{table}
Before reporting the case study, in which data gathering and analysis was used to support the development and running of a final-year module taken by an entire year group in a UK University Computer Science Department over three years, the next section will frame the research questions that the paper addresses. 

\section{Research Questions}

Two sets of research questions – those related to the study itself, and those seeking to understand the broader implications of the study – were framed. At the start of the study, the second set were less important to us, but with the rapid spread of on-line teaching methods, understanding the potential and limitations of specific studies for more general application, now seems critical.

The \textit{post hoc} analysis sought to address four research questions: 

RQ1: How well can we pinpoint students in danger of failing?

RQ2: Is engagement linked to final achievement? If so, can we predict students’ final grades from the way they approached the course?

RQ3: How does failing the coursework first time impact the final outcome?

RQ4: How stable are any identified patterns over time (i.e., are patterns stable between cohorts; are there improvements in specific measures over time)?

The wider research questions are:

RQ5: To what extent could this trial be replicated elsewhere?

RQ6: How generalisable are the findings likely to have been?

It is important to note that this study is different from many others in the area since it is concerned with final-year students, therefore a number of students (typically around 18$\%$ of the initial number who began the programme in the first year) have already been removed from the cohort, either through academic failure or through electing not to continue.

\section{The Study Context: background to the module, a timeline of the changes made, and rationale for the changes}

From 2010/11, a 20-credit module on Software Project Management, compulsory to all final-year (120-credit) students (145-181 per cohort) in the Department of Computer Science, Brunel University London, was progressively redesigned away from a lectures-based course with single exam assessment to one with significant on-line study, face-to-face engagement, and a two-stage assessment in which students who passed the coursework passed the entire module (at the threshold pass level) and went on to sit a grade-threshold-based exam for higher grades.

The framework of enquiry was one in which changes were proposed through the usual departmental quality assurance/change management systems and evaluated in-year using a range of sources, from attendance and on-line activity data, to questionnaires and quizzes, including the departmental and university systems for collecting student views. The normal panels and boards of examiners reviewed the outcomes. In terms of approach, this was an exploratory research study.

\begin{figure}[!htbp]
\includegraphics[width=\textwidth]{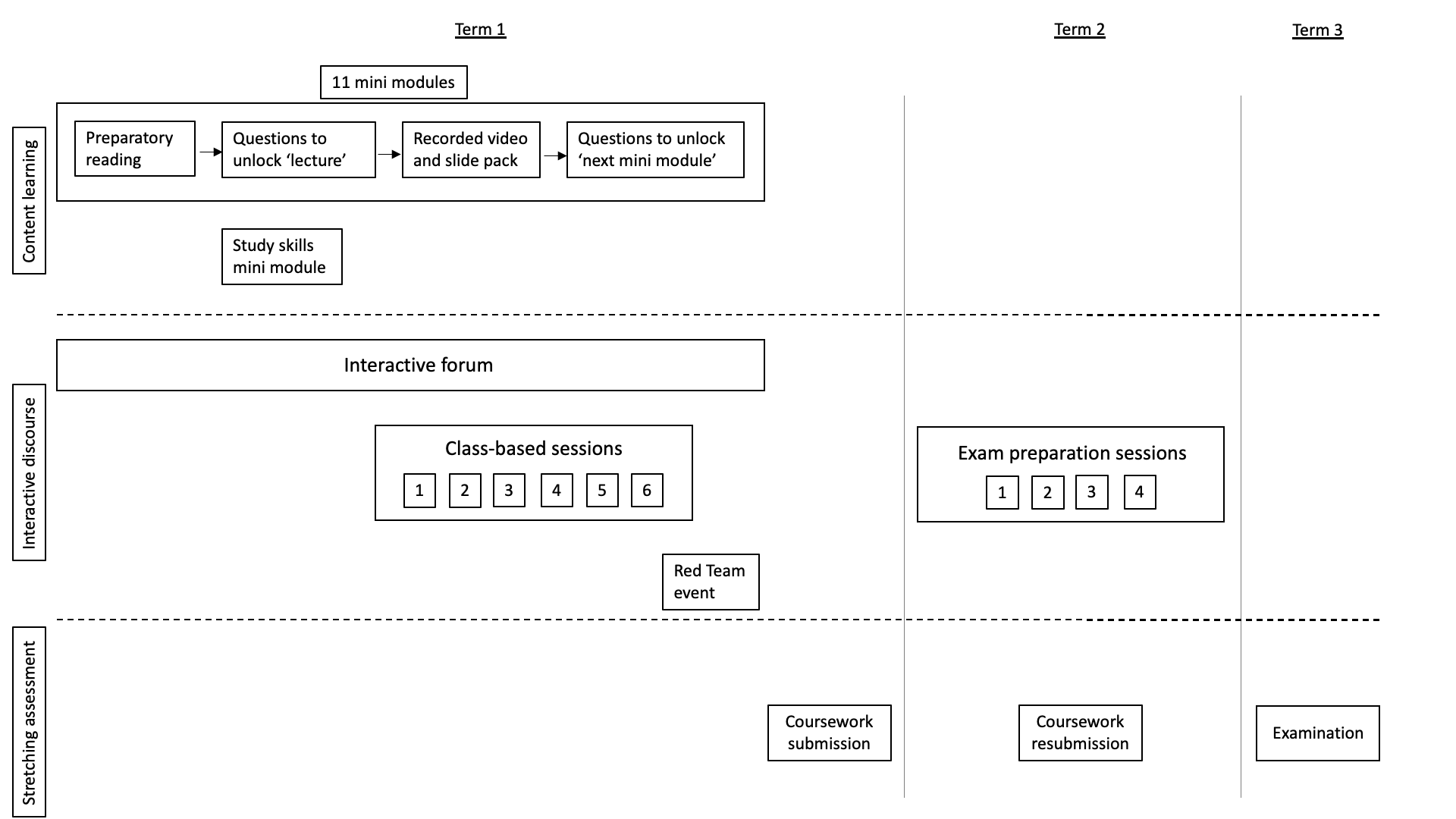}
\caption{Overall structure of the module organised by term and showing the strands of the pedagogic model used}
\label{fig:overall_structure_module_organised_term}
\end{figure}

The pedagogic model that underpinned the module being studied had three components – content learning, interactive discourse and stretching assessment – although we are aware of more sophisticated models, especially in the context of learning technology \cite{holmes2019learning} \cite{laurillard2009the} \cite{laurillard2013rethinking}.The module was set up as a repeated sequence of content learning, discourse and assessment (see Figure \ref{fig:overall_structure_module_organised_term}). At the end of the first term, the assessment consisted of producing a project bid within a structured framework of eight documents, and at the end of the second term, a threshold-based written exam. Those passing the coursework gained a pass for the module as a whole (at the lowest passing grade, D-) and went on at the start of the third term to sit the exam (which determined the final grade between D- and A$\ast$). Those who failed the coursework had an opportunity to be reassessed in Term 2 and one of the questions explored is whether those who were reassessed in the coursework could do as well as those who passed first time. 

In order to make changes within the constraints of a normal academic cycle, we aimed to make as much use as we could of existing subject-based material within the module. This had the effect of changing the delivery radically while maintaining continuity of the module’s content and learning outcomes.

The context learning was based upon the standard, hour-long scheduled lectures as delivered in previous years (filmed from the back of the lecture theatre so that students were not identifiable). These were edited to remove such elements as instructions being issued to a specific cohort (or the occasional vigorous exchange between lecturer and student body). Post-editing, most of the lectures were around 35-40 minutes in length, and students had access to the original slide sets.

At the time, the VLE did not offer the capability to capture information around how students engaged with this material, over and above that they had engaged and at what time. This was how engagement was tracked across the cohort (though, for pragmatic reasons, we did undertake analysis based on the time of engagement). The lectures themselves had followed the course textbook and so the associated reading/preparatory material was set up to be done before the video of the relevant lecture. In practice, a short quiz was used to check that students had read the material and a second quiz was used to ensure that they had watched the video. The submission/timestamp of the quiz results gave the best estimate of when the reading and lectures were being studied by each student. 

The quizzes were designed to check for engagement, not understanding, and we accepted that some students engaged at this stage in a perfunctory manner. However, the results also showed that this was certainly not the trend. A week’s ‘lecture’ therefore became a \textit{mini module} consisting of a reading and a quiz that opened up the video and slide pack, followed by a quiz that opened up the start of the next \textit{mini module}. Eleven of the 12 Term 1 mini modules took this form, with one being a study skills lecture from a librarian on how to access relevant sources, find references, and so forth. In Term 2, students were expected to structure their own learning and so lectures and videos only were available (i.e., there were no quizzes and so no ‘unlocking’ of materials); as such engagement was not tracked in the same way. The aim was for the students to complete the on-line element in four weeks at the start of Term 1.

In Term 1, a set of six class-based sessions were the core interactive element intended to bridge between the content and its practical application. The sessions were couched around a set of structured documents that the students needed to complete as their coursework submission – a project bid in response to a brief. The documents were intended to take concepts that the cohort had encountered on-line – such as work breakdown structures – and enable the students to use them to build their individual bids. At their best, the interactive class session enabled students to explain to the rest of the class how they were tackling a given document and to discuss the approaches taken. This part was hardest to implement because we chose not to use graduate teaching assistants (GTAs) to run these sessions and so it was hard to break the class into more than three groups (of typically 50-65 students each) because of the workload constraints/commitments of the academics leading the sessions. Most of the core learning took place in Term 1.

In Term 2, further interactive sessions were used, which discussed the exam scenario and allowed a set of trial exam questions to be explored (three exams papers were written early on in the cycle: one for the main summer exam period, one for any resit required, and one for the class to use for their own learning in and outside of the interactive sessions).

Other interactive elements used in the module were an introductory afternoon with a construction game to kick off the module at the start of Term 1 (focussed on project roles and communication), an on-line forum for shared questions and answers, and an event towards the end of Term 1 in which students could bring their (part- or un-finished) coursework to receive peer review and formative feedback from staff. This latter idea was modified from an industrial idea where bids were given a thorough review by any interested party before submission. In the form one of us had encountered it, it was called a Red Team event and the proposal would be pinned on the walls of a review room, and anyone with any comment at all would be encouraged to write in red on or near the relevant page. 

The coursework element of the assessment was to produce a bid for a software project using eight documents that were loosely specified. It was open-ended in that students had to scope the project and cost it using techniques covered in the teaching/learning materials, but they were also required to research and identify suitable overhead and contingency rates, salary levels, and to find comparator projects to calibrate their own calculations. The exam, scenario/case based, with the case circulated well in advance of the examination, was three hours in duration and comprised questions set to correspond to the grade thresholds (a C-grade question, a B-grade question, and an A-grade question).

Data on student attendance was collected using a purpose-built system (based on scanning student ID cards) and the VLE provided quiz scores and information on when each student sat the quizzes. 

By 2014/15, the module had reached a fairly stable stage and the first year of data was collected at this time. In general terms, the timings of teaching and learning activities were as follows:

\textbf{Weeks 1-4}. Introductory session, one lecture and a seminar. When it became clear within the first two weeks of Term 1 that the students disliked the reading, a speed-reading course was laid on at short notice (in week 4) and students registered on-line. Weekly surgeries were run to support students in their learning (and some students were asked to present at a surgery, where their lack of progress/engagement with the module material warranted it).

This time was used to ‘cleanse’ the central university student records database and work out who was really on the module. Meanwhile, it became clear that a significant number of students were going to complete the on-line element in three weeks, and so preparations were made for one seminar set (comprising these students) to start the interactive seminars a week early, as a fast-track.

\textbf{Weeks 5-12} (remainder of Term 1). The coursework was released, the interactive seminars run and, at the very end of Term 1, a Red Team event (described above) was run in week 12 for peer review and formative feedback.

\textbf{Term 2}. Coursework was handed in at the start of Term 2. A mix of 2-3 lectures and seven on-line lectures were available (including guest lectures). The cohort watched a movie together (Apollo 13) to support idea of managing away from failure.

Interactive seminars to prepare for the exam were run in the second half of Term 2.

\textbf{Term 3}. Exam

For 2015/16, the following changes were implemented:

\textbf{Weeks 1-4}. The speed reading course was scheduled from the outset of this run of the module. The fast-track system was retained. However, the coursework was released in week 3 to allow students to start it earlier.

\textbf{Weeks 5-12} (remainder of Term 1). In week 6 students were asked to submit their attempts for two foundational documents (as a ‘stage gate’), since it was realised that failure to set up their bid properly had a domino effect on achievement in the overall coursework. Formative feedback was given on the two documents and the results stored as summative outcomes for those who passed those documents. Coursework was handed in at the end Term 1 and so the Red Team event was moved to week 10.

\textbf{Term 2}. No further significant changes were made from 2014/15.

\textbf{Term 3.} No significant changes were made from 2014/15.

For 2016/17, the following changes were implemented:

\textbf{Weeks 1-4}. The early assessment of the key documents was moved one week earlier, to week 5, and the coursework was released on the first day of Term 1.

\textbf{Weeks 5-12} (remainder of Term 1). The Red Team event was moved to week 9 and the coursework was handed in during week 10. The fast-track seminars were dropped – all cohorts ran at the same time.

\textbf{Term 2}. No further significant changes were made from 2015/16.

\textbf{Term 3.} No significant changes from 2015/16.

A summary of the changes to the learning and teaching activities across the three years is presented in Table 2. 

\begin{table}[!htbp]
\renewcommand{\arraystretch}{1.3}
\begin{adjustbox}{max width=\textwidth}
\begin{tabular}{p{1.42cm}p{5.49cm}p{5.28cm}p{4.88cm}p{0.6cm}p{1.63cm}}
\hhline{~-----}
\multicolumn{1}{p{1.42cm}}{\multirow{2}{*}{\parbox{1.42cm}{}}} & 
\multicolumn{3}{|p{15.65cm}}{\textbf{Teaching and Learning Activities }} & 
\multicolumn{1}{p{0.6cm}}{} & 
\multicolumn{1}{p{1.63cm}|}{} \\ 
\hhline{~-----}
\multicolumn{1}{p{1.42cm}}{} & 
\multicolumn{2}{|p{10.77cm}}{\textbf{Term 1 }} & 
\multicolumn{1}{|p{4.88cm}}{\multirow{2}{*}{\parbox{4.88cm}{\textbf{Term 2 }}}} & 
\multicolumn{1}{|p{0.6cm}}{\multirow{2}{*}{\parbox{0.6cm}{}}} & 
\multicolumn{1}{p{1.63cm}|}{\multirow{2}{*}{\parbox{1.63cm}{\textbf{Term 3 }}}} \\ 
\hhline{---~~~}
\multicolumn{1}{|p{1.42cm}}{\textbf{Year }} & 
\multicolumn{1}{|p{5.49cm}}{\textbf{Weeks 1-4 }} & 
\multicolumn{1}{|p{5.28cm}}{\textbf{Weeks 5-12 }} & 
\multicolumn{1}{|p{4.88cm}}{} & 
\multicolumn{1}{|p{0.6cm}}{} & 
\multicolumn{1}{p{1.63cm}|}{} \\ 
\hline
\multicolumn{1}{|p{1.42cm}}{2014/15 } & 
\multicolumn{1}{|p{5.49cm}}{\begin{itemize}
	\item Access to online ‘mini modules’ \newline
	\item Introductory session \newline
	\item One lecture and a seminar \newline
	\item Short-notice speed-reading course \newline
	\item Surgeries weeks 2 $\&$ 4 \newline
	\item Fast-track seminar group introduced (starting in week 4) to cater for those finishing ‘mini modules’ early \end{itemize}
} & 
\multicolumn{1}{|p{5.28cm}}{\begin{itemize}
	\item Coursework release (week 5) \newline
	\item Interactive seminars (weeks 5-8) \newline
	\item Red Team event held in week 12 \newline
	\item Surgeries weeks 10 $\&$ 11\end{itemize}
} & 
\multicolumn{1}{|p{4.88cm}}{\begin{itemize}
	\item Coursework submission (start of Term 2) \newline
	\item 2-3 face-to-face lectures \newline
	\item 7 online lectures \newline
	\item Movie (week 23) \newline
	\item Interactive seminars – exam preparation (weeks 25-27)\end{itemize}
} & 
\multicolumn{1}{|p{0.6cm}}{} & 
\multicolumn{1}{p{1.63cm}|}{Exam } \\ 
\hline
\multicolumn{1}{|p{1.42cm}}{2015/16 } & 
\multicolumn{1}{|p{5.49cm}}{As for 2014/15 but: \newline
\begin{itemize}
	\item Speed reading course schedule from outset \newline
	\item Fast-track seminar group retained\end{itemize}
} & 
\multicolumn{1}{|p{5.28cm}}{As for 2014/15 but: \newline
\begin{itemize}
	\item Coursework released earlier (week 3) \newline
	\item Students asked to submit two foundational documents from coursework for formative feedback (stage gate) (week 6) \newline
	\item Red Team event moved earlier (week 10) \newline
	\item Coursework submission moved to end of Term 1 (week 12)\end{itemize}
} & 
\multicolumn{1}{|p{4.88cm}}{No further significant changes made from 2014/15} & 
\multicolumn{1}{|p{0.6cm}}{} & 
\multicolumn{1}{p{1.63cm}|}{Exam } \\ 
\hline
\multicolumn{1}{|p{1.42cm}}{2016/17 } & 
\multicolumn{1}{|p{5.49cm}}{Coursework released on first day of Term 1. Submission of two foundational documents from coursework for formative feedback (stage gate) moving one week earlier (week 5)} & 
\multicolumn{1}{|p{5.28cm}}{\begin{itemize}
	\item Red Team event moved earlier (week 9) \newline
	\item Coursework submission moved two weeks earlier (week 10) \newline
	\item The fast-track seminar group was dropped; all groups started in week 5\end{itemize}
} & 
\multicolumn{1}{|p{4.88cm}}{No further significant changes made from 2015/16} & 
\multicolumn{1}{|p{0.6cm}}{} & 
\multicolumn{1}{p{1.63cm}|}{Exam } \\ 
\hline
\end{tabular}
\end{adjustbox}
\caption{A summary of the changes to the learning and teaching activities 2014/15-2016/17 12}
\label{tab:summary_changes_learning_and_teaching}\end{table}
\section{Data Collection and Cleaning}

Having described the context of the module, this section explains the sources from which the structured data (quantitative and categorical) were collected on which the \textit{post hoc} analysis was based). We also consider the practical difficulties that were faced in analysing the data across the three years, and the features and value ranges within the dataset.

At the end of each year, a series of spreadsheets existed, showing cohort engagement (quiz scores from on-line tests and presence at seminars, surgeries and other events), and another set with coursework scores and exam results. This was a rather disparate set of documents from which to start working. There was some overlap between the module-level, departmental, and university records.

Three of the authors then collected this raw data and imported it into a new set of spreadsheets, resulting in datasets from the three student cohorts (2014/15, 2015/16 and 2016/17), totalling data for 492 students. As noted, the original data had come from a range of sources including: (i) the VLE (ii) the university’s student record system, SITS; (iii) official documents submitted to, and minutes of, Panel and Board of Examiners’ meetings; and (iv) local data files collected informally by the module team\footnote{ Our sanitised data and R code is made publicly available at https://figshare.com/articles/dataset/\_/12816203 and https://figshare.com/articles/software/\_/12816185 respectively.}.

The creation of the datasets, and preparing them for analysis, presented a number of difficulties:

\begin{enumerate}
	\item The data needed to be matched across sources.

	\item The data collected, and the labels, were not consistent between years.

	\item Decisions about which students to include were not necessarily consistent (e.g., if a student 'dropped out' during the year could this count as a fail or a withdrawal).

	\item The complexity of data available, particularly from the VLE, was overwhelming\footnote{ For example, the VLE kept data on every single page visited by each student, along with time and date}, which led to the need to aggregate and simplify.

	\item Values were inconsistent (e.g., with leading or trailing spaces and inconsistent capitalisation).

	\item The delivery schedules (as noted above) varied between cohorts, so workshops might be differently timed or organised from year to year.

\end{enumerate}
Rationalising and cleaning the data became a major task that required about 50 hours for one of the authors. From this experience we would strongly recommend making data collection a prospective exercise. 

The number of students in each of year cohorts is given in Table 3.

\begin{table}[!htbp]
\renewcommand{\arraystretch}{1.3}
\begin{adjustbox}{max width=\textwidth}
\begin{tabular}{p{2.25cm}p{3.47cm}}
\hline
\multicolumn{1}{|p{2.25cm}}{\textbf{Cohort}} & 
\multicolumn{1}{|p{3.47cm}|}{\textbf{Student numbers}} \\ 
\hline
\multicolumn{1}{|p{2.25cm}}{2014\_15 } & 
\multicolumn{1}{|p{3.47cm}|}{181} \\ 
\hline
\multicolumn{1}{|p{2.25cm}}{2015\_16 } & 
\multicolumn{1}{|p{3.47cm}|}{166} \\ 
\hline
\multicolumn{1}{|p{2.25cm}}{2016\_17 } & 
\multicolumn{1}{|p{3.47cm}|}{145} \\ 
\hline
\end{tabular}
\end{adjustbox}
\caption{Number of students in each of the year cohorts}
\label{tab:features_that_were_collected_and}\end{table}

The features that were collected, and their value ranges, are summarised in Table 4.

\begin{table}[!htbp]
\renewcommand{\arraystretch}{1.3}
\begin{adjustbox}{max width=\textwidth}
\begin{tabular}{p{3.12cm}p{3.06cm}p{9.53cm}}
\hline
\multicolumn{1}{|p{3.12cm}}{\textbf{Feature}} & 
\multicolumn{1}{|p{3.06cm}}{\textbf{Values}} & 
\multicolumn{1}{|p{9.53cm}|}{\textbf{Explanation}} \\ 
\hline
\multicolumn{1}{|p{3.12cm}}{Year} & 
\multicolumn{1}{|p{3.06cm}}{2014-15, 2015-16, 2016-17} & 
\multicolumn{1}{|p{9.53cm}|}{Year of cohort} \\ 
\hline
\multicolumn{1}{|p{3.12cm}}{ID} & 
\multicolumn{1}{|p{3.06cm}}{} & 
\multicolumn{1}{|p{9.53cm}|}{This was only used for cross-checking purposes and all data in the analysis is anonymised} \\ 
\hline
\multicolumn{1}{|p{3.12cm}}{OverallGrade} & 
\multicolumn{1}{|p{3.06cm}}{A$\ast$, ... , F} & 
\multicolumn{1}{|p{9.53cm}|}{17 module grades used by Brunel University London} \\ 
\hline
\multicolumn{1}{|p{3.12cm}}{CalculatedMark} & 
\multicolumn{1}{|p{3.06cm}}{17, ... ,1} & 
\multicolumn{1}{|p{9.53cm}|}{Numeric equivalent of grade} \\ 
\hline
\multicolumn{1}{|p{3.12cm}}{OverallPass} & 
\multicolumn{1}{|p{3.06cm}}{T or F} & 
\multicolumn{1}{|p{9.53cm}|}{Did the student pass the module (i.e., obtain at least D-)?} \\ 
\hline
\multicolumn{1}{|p{3.12cm}}{CW1\_Pass} & 
\multicolumn{1}{|p{3.06cm}}{T or F} & 
\multicolumn{1}{|p{9.53cm}|}{Did the student pass the coursework first time?} \\ 
\hline
\multicolumn{1}{|p{3.12cm}}{RCW\_Pass} & 
\multicolumn{1}{|p{3.06cm}}{T or F or NA} & 
\multicolumn{1}{|p{9.53cm}|}{Did the student pass the coursework on Term 2 reassessment?} \\ 
\hline
\multicolumn{1}{|p{3.12cm}}{W1\_3Engt} & 
\multicolumn{1}{|p{3.06cm}}{[0, 1]} & 
\multicolumn{1}{|p{9.53cm}|}{Proportion of physical events (weeks 1-3) attended (e.g., workshops, etc.)} \\ 
\hline
\multicolumn{1}{|p{3.12cm}}{W4\_12Engt} & 
\multicolumn{1}{|p{3.06cm}}{[0, 1]} & 
\multicolumn{1}{|p{9.53cm}|}{Proportion of physical events (weeks 4-12) attended} \\ 
\hline
\multicolumn{1}{|p{3.12cm}}{W16\_24Engt} & 
\multicolumn{1}{|p{3.06cm}}{[0, 1]} & 
\multicolumn{1}{|p{9.53cm}|}{Proportion of physical events (weeks 16-24) attended} \\ 
\hline
\multicolumn{1}{|p{3.12cm}}{W1\_3OnL\_Engmt} & 
\multicolumn{1}{|p{3.06cm}}{[0, 1]} & 
\multicolumn{1}{|p{9.53cm}|}{Proportion of online activities (weeks 1-3) undertaken (e.g., online quizzes, supporting reading, etc.)} \\ 
\hline
\end{tabular}
\end{adjustbox}
\caption{The features that were collected and their value ranges}
\end{table}

It is important to note that all of the engagement features were normalised because there were differing numbers of events between years.

The OverallPass rate across the three years is: 467/492, $\sim$95$\%$. In total there were 25 module failures, making failure (fortunately) a rare event. In part, this is a consequence of the module in question being in the final-year where module failure rates are generally lower that in earlier stages of the degree. It does, however, have implications from a prediction perspective as this means that there is a considerable shortage of failure training instances (i.e., the data are imbalanced).

All cases are complete with the exception of RCW\_Pass, which is only relevant if the student has previously failed their first attempt at the coursework (i.e., CW1\_Pass $=$ F)

\section{Exploratory Analysis of the Data}

Having explained the features and their value ranges, this section will present the exploratory analysis of the data, beginning with a comparison of the year-based cohorts in terms of overall pass rate, grade frequencies and engagement, before addressing research questions 1-4 (RQ1-RQ4). 

Table 5 presents the module pass rate by year. Although the overall fail rate is low, in 2015-16 it seems a little anomalous, with a much lower level than for the other two cohorts (see Table 5). 

\begin{table}[!htbp]
\renewcommand{\arraystretch}{1.3}
\begin{adjustbox}{max width=\textwidth}
\begin{tabular}{p{2.65cm}p{1.9cm}p{1.88cm}p{1.85cm}}
\hline
\multicolumn{1}{|p{2.65cm}}{\textbf{OverallPass}} & 
\multicolumn{1}{|p{1.9cm}}{\textbf{2014\_15}} & 
\multicolumn{1}{|p{1.88cm}}{\textbf{2015\_16}} & 
\multicolumn{1}{|p{1.85cm}|}{\textbf{2016\_17}} \\ 
\hline
\multicolumn{1}{|p{2.65cm}}{Fail} & 
\multicolumn{1}{|p{1.9cm}}{14} & 
\multicolumn{1}{|p{1.88cm}}{1} & 
\multicolumn{1}{|p{1.85cm}|}{10} \\ 
\hline
\multicolumn{1}{|p{2.65cm}}{Pass} & 
\multicolumn{1}{|p{1.9cm}}{167} & 
\multicolumn{1}{|p{1.88cm}}{165} & 
\multicolumn{1}{|p{1.85cm}|}{135} \\ 
\hline
\end{tabular}
\end{adjustbox}
\caption{Module pass rate by year}
\end{table}

\begin{figure}[!htbp]
\includegraphics[width=13.76cm,height=7.66cm]{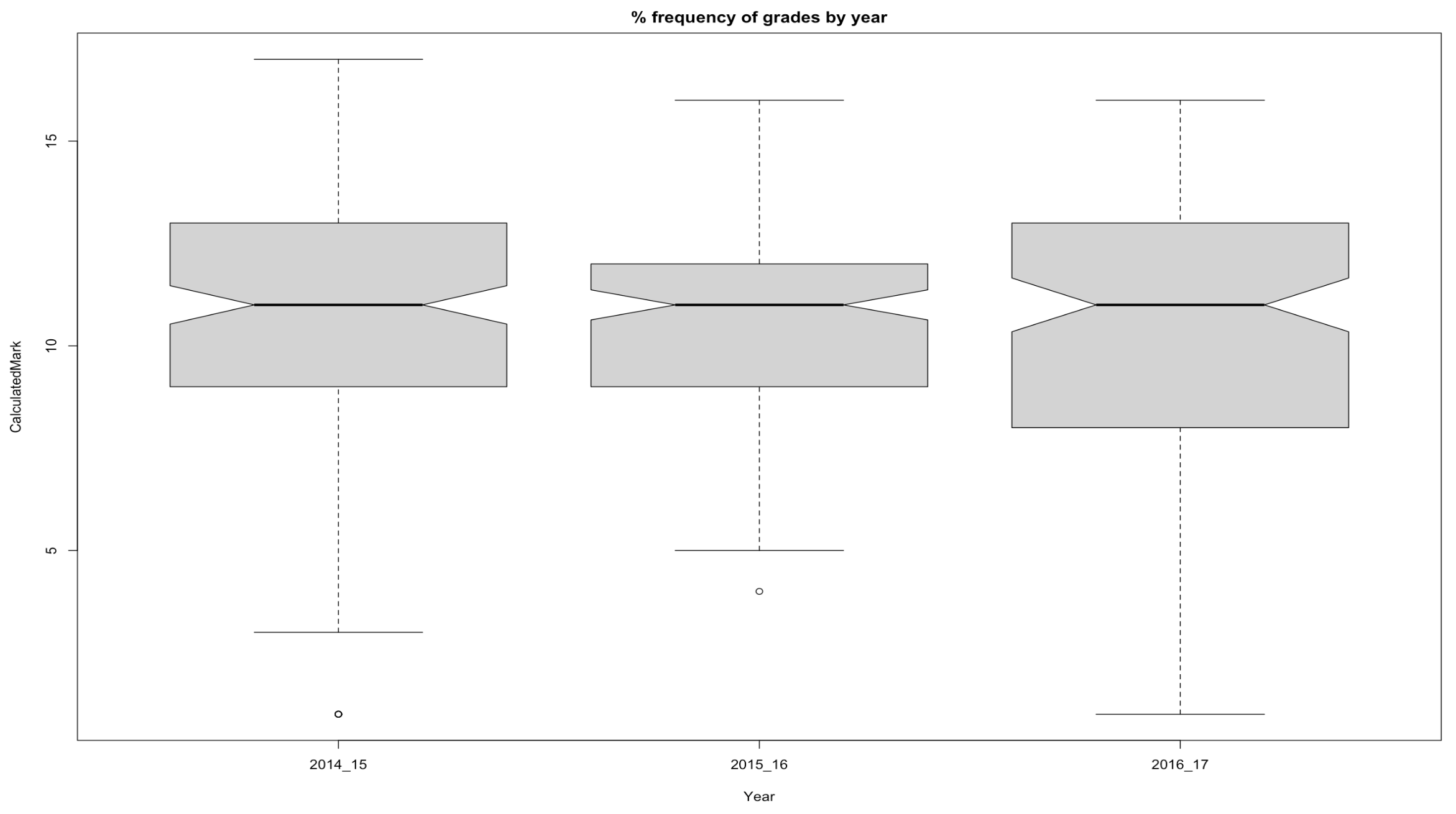}
\caption{Side by side boxplots showing the distribution of grades (A$\ast$ $=$ 17, ... , F $=$1) and the notches show the 95$\%$ CIs}
\label{fig:side_side_boxplots_showing_distribution}
\end{figure}

Breaking this down, the distribution of grade by year as side-by-side boxplots were analysed (see Figure \ref{fig:side_side_boxplots_showing_distribution}). What is interesting is that the median grade, denoted by the horizontal line, is constant at 11 $=$ B- for all three years. As can be seen, though, there is a good deal more variability for 2014/15 and 2016/17.

\begin{figure}[!htbp]
\includegraphics[width=10.17cm,height=7.23cm]{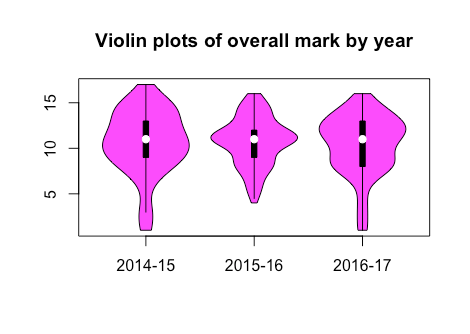}
\caption{Side by side violin plots showing the distribution of grades (A$\ast$ $=$ 17, ... , F $=$1)}
\label{fig:side_side_violin_plots_showing}
\end{figure}

Alternatively, the distribution of grades can be presented using a kernel density function as side-by-side violin plots (see Figure \ref{fig:side_side_violin_plots_showing}).

From these curves, there is nothing to support the traditional (and, to our minds, somewhat unfounded) idea of grades following a bell or normal curve \cite{fashing1981the} \cite{gasevic2016learning}. Probably the most interesting additional finding is that there is some evidence of multi-modality: at least two sub-distributions of performance are visible in the 2014-15 and 2016-17 plots, while three sub-distributions are conceivably behind the shape of the 2015-16 plot.

 This fits with the experience of those teaching on the module, who usually encountered a group of keen students working exceptionally hard and often worrying that they might fail, and a group of students that did very little until halfway through Term 1 or later. There was often a middle-group made up of students who moved slowly to begin with but caught up as the year progressed. We believe this phenomenon merits further investigation, especially since it is not clear that different cohorts even approached the module in the same way.

\subsection{Relationships between variables}

\begin{figure}[!htbp]
\includegraphics[width=\textwidth]{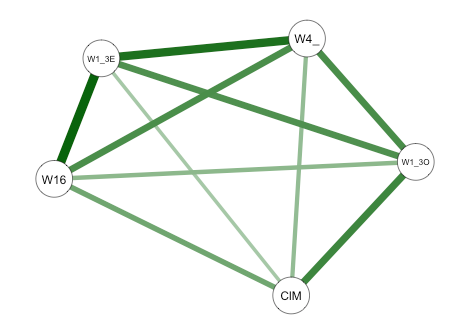}
\caption{Correlogram of the student engagement factors and OverallMark (W1\_3E $=$ W1-3 engagement, W4\_ $=$ W4-12 engagement, W16 $=$ W16-24 engagement, W1\_3O $=$ W1-3 online engagement and ClM $=$ calculated overall mark)}
\label{fig:correlogram_student_engagement_factors_and}
\end{figure}

Having compared the cohorts, this section considers the relationships between the five numeric features that that were collected (see Table \ref{tab:kendalls_tau_cross_correlations}). Figure \ref{fig:correlogram_student_engagement_factors_and} shows the Kendall's Tau coefficients (chosen as a robust correlation method) between the features, with green indicating positive and red negative correlations, and the width of the line being proportional to the strength. The coefficients are also given in Table 6. From this, it would appear that all four types of engagement are positively associated with the final module mark, but that the strongest association is with early online engagement.
\begin{table}[!htbp]
\renewcommand{\arraystretch}{1.3}
\begin{adjustbox}{max width=\textwidth}
\begin{tabular}{p{3.39cm}p{2.11cm}p{2.13cm}p{2.34cm}p{2.54cm}p{3.39cm}}
\hhline{~-----}
\multicolumn{1}{p{3.39cm}}{} & 
\multicolumn{1}{|p{2.11cm}}{Calculated Mark} & 
\multicolumn{1}{|p{2.13cm}}{W1\_3Engt} & 
\multicolumn{1}{|p{2.34cm}}{W4\_12Engt} & 
\multicolumn{1}{|p{2.54cm}}{W16\_24Engt} & 
\multicolumn{1}{|p{3.39cm}|}{W1\_3OnL\_Engmt} \\ 
\hline
\multicolumn{1}{|p{3.39cm}}{Calculated Mark} & 
\multicolumn{1}{|p{2.11cm}}{1.000} & 
\multicolumn{1}{|p{2.13cm}}{0.118} & 
\multicolumn{1}{|p{2.34cm}}{0.146} & 
\multicolumn{1}{|p{2.54cm}}{0.193} & 
\multicolumn{1}{|p{3.39cm}|}{0.263} \\ 
\hline
\multicolumn{1}{|p{3.39cm}}{W1\_3Engt} & 
\multicolumn{1}{|p{2.11cm}}{0.118} & 
\multicolumn{1}{|p{2.13cm}}{1.000} & 
\multicolumn{1}{|p{2.34cm}}{0.310} & 
\multicolumn{1}{|p{2.54cm}}{0.341} & 
\multicolumn{1}{|p{3.39cm}|}{0.239} \\ 
\hline
\multicolumn{1}{|p{3.39cm}}{W4\_12Engt} & 
\multicolumn{1}{|p{2.11cm}}{0.146} & 
\multicolumn{1}{|p{2.13cm}}{0.310} & 
\multicolumn{1}{|p{2.34cm}}{1.000} & 
\multicolumn{1}{|p{2.54cm}}{0.246} & 
\multicolumn{1}{|p{3.39cm}|}{0.248} \\ 
\hline
\multicolumn{1}{|p{3.39cm}}{W16\_24Engt} & 
\multicolumn{1}{|p{2.11cm}}{0.193} & 
\multicolumn{1}{|p{2.13cm}}{0.341} & 
\multicolumn{1}{|p{2.34cm}}{0.246} & 
\multicolumn{1}{|p{2.54cm}}{1.000} & 
\multicolumn{1}{|p{3.39cm}|}{0.155} \\ 
\hline
\multicolumn{1}{|p{3.39cm}}{W1\_3OnL\_Engmt} & 
\multicolumn{1}{|p{2.11cm}}{0.263} & 
\multicolumn{1}{|p{2.13cm}}{0.239} & 
\multicolumn{1}{|p{2.34cm}}{0.248} & 
\multicolumn{1}{|p{2.54cm}}{0.155} & 
\multicolumn{1}{|p{3.39cm}|}{1.000} \\ 
\hline
\end{tabular}
\end{adjustbox}
\caption{Kendall's Tau cross correlations}
\label{tab:kendalls_tau_cross_correlations}\end{table}

This result is very interesting and enables us to see something of the success of a core principle in redesigning the course, namely to move the ‘book learning’ early in the module and to encourage the students to undertake this largely on their own from the outset.

\section{Findings}

Having presented some basic descriptive analysis of the student data, we now address our four research questions. For RQ1-3 we pool all three cohorts but for RQ4 we disaggregate.

\textbf{RQ1: How well can we pinpoint students in danger of failing?}

Given the correlation analysis between the five numeric features, this section explores some predictive models. To this end, we considered two approaches: (i) logistic regression models to determine whether a student passed or not (since the dependent variable is dichotomous); and (ii) ordinary least squares multiple regression to model the final Calculated Mark which is a score (1 to 17) representing the Overall Grade. 

Our simple logistic model to try to predict student success or failure in a dichotomous fashion uses all four engagement factors and combines data from all three cohorts. We focus on engagement since we hope that this is something that can be manipulated (or, at least, it is a factor that we can try to influence). This yields the model shown in Table 6.  

We need to consider several aspects of the model. First, the goodness of fit. Although not an exact analogue to the conventional adjusted R\textsuperscript{2}, we use McFadden's R\textsuperscript{2} index \cite{smith2013a}. Second, we need to consider issues of multi-collinearity since we have already identified some cross correlations (see Figure \ref{fig:correlogram_student_engagement_factors_and}). For this we examine the variance inflation (VIF) scores, which ideally should be less than 5.

 We commence with a naïve model (LR model 1) where we enter all possible predictor variables that would be available prior to the final module outcome (see Table 7).

\begin{table}[!htbp]
\renewcommand{\arraystretch}{1.3}
\begin{adjustbox}{max width=\textwidth}
\begin{tabular}{p{3.15cm}p{2.29cm}p{2.29cm}p{2.29cm}}
\hhline{~---}
\multicolumn{1}{p{3.07cm}}{} & 
\multicolumn{1}{|p{2.29cm}}{\textbf{Estimate}} & 
\multicolumn{1}{|p{2.29cm}}{\textbf{Std. Error}} & 
\multicolumn{1}{|p{2.29cm}|}{\textbf{p}} \\ 
\hline
\multicolumn{1}{|p{3.07cm}}{(Intercept)} & 
\multicolumn{1}{|p{2.29cm}}{0.819} & 
\multicolumn{1}{|p{2.29cm}}{0.427} & 
\multicolumn{1}{|p{2.29cm}|}{0.055} \\ 
\hline
\multicolumn{1}{|p{3.07cm}}{W1\_3Engt} & 
\multicolumn{1}{|p{2.29cm}}{1.736} & 
\multicolumn{1}{|p{2.29cm}}{1.018} & 
\multicolumn{1}{|p{2.29cm}|}{0.088} \\ 
\hline
\multicolumn{1}{|p{3.07cm}}{W4\_12Engt} & 
\multicolumn{1}{|p{2.29cm}}{0.394} & 
\multicolumn{1}{|p{2.29cm}}{0.948} & 
\multicolumn{1}{|p{2.29cm}|}{0.678} \\ 
\hline
\multicolumn{1}{|p{3.07cm}}{W16\_24Engt} & 
\multicolumn{1}{|p{2.29cm}}{1.534} & 
\multicolumn{1}{|p{2.29cm}}{1.064} & 
\multicolumn{1}{|p{2.29cm}|}{0.150} \\ 
\hline
\multicolumn{1}{|p{3.07cm}}{W1\_3OnL\_Engmt} & 
\multicolumn{1}{|p{2.29cm}}{2.681} & 
\multicolumn{1}{|p{2.29cm}}{1.010} & 
\multicolumn{1}{|p{2.29cm}|}{0.008} \\ 
\hline
\end{tabular}
\end{adjustbox}
\caption{Logistic regression model to predict module passing students}
\label{tab:logistic_regression_model_predict_module}\end{table}
This reveals W1-3 (weeks 1-3) online engagement to be the most important factor in the model (because the beta coefficient is greatest (and has the smallest p value). This is followed by W1-3 (face to face) engagement. Thus, we see some evidence that early on-line and face-to-face engagement (i.e., in the first three weeks of the year-long module) are the strongest predictor variables in predicting whether a student is likely to pass or fail. However, McFadden's R\textsuperscript{2} index is 0.148 suggesting low explanatory value. The VIF analysis yields all predictor variables are <2 suggesting no particular problems with multicollinearity. Since the model does not have good fit we do not proceed in this direction.

The other approach to modelling is to use standard multiple regression and Grade as the dependent variable. We used a stepwise procedure based on Akaike information criterion (AIC) and both forward and backward selection to try to determine the best subset of predictors to include. This led to the model shown in Table 8. This again highlights W1-3 online engagement as being the most important predictor. Again the fit is poor (the Adjusted R-squared $=$ 0.161). As with the logistic regression, VIF analysis yields all predictor variables are <2 suggesting no particular problems with multicollinearity.

\begin{table}[!htbp]
\renewcommand{\arraystretch}{1.3}
\begin{adjustbox}{max width=\textwidth}
\begin{tabular}{p{3.98cm}p{3.98cm}p{3.98cm}p{3.98cm}}
\hhline{~---}
\multicolumn{1}{p{3.98cm}}{} & 
\multicolumn{1}{|p{3.98cm}}{Estimate} & 
\multicolumn{1}{|p{3.98cm}}{Std. Error} & 
\multicolumn{1}{|p{3.98cm}|}{Pr(>$\vert$t$\vert$)} \\ 
\hline
\multicolumn{1}{|p{3.98cm}}{(Intercept)} & 
\multicolumn{1}{|p{3.98cm}}{8.1537} & 
\multicolumn{1}{|p{3.98cm}}{0.3217} & 
\multicolumn{1}{|p{3.98cm}|}{< 2e-16} \\ 
\hline
\multicolumn{1}{|p{3.98cm}}{W1\_3OnL\_Engmt} & 
\multicolumn{1}{|p{3.98cm}}{3.1685} & 
\multicolumn{1}{|p{3.98cm}}{0.4299} & 
\multicolumn{1}{|p{3.98cm}|}{7.34E-13} \\ 
\hline
\multicolumn{1}{|p{3.98cm}}{W4\_12Engt} & 
\multicolumn{1}{|p{3.98cm}}{0.8239} & 
\multicolumn{1}{|p{3.98cm}}{0.5612} & 
\multicolumn{1}{|p{3.98cm}|}{0.142729} \\ 
\hline
\multicolumn{1}{|p{3.98cm}}{W16\_24Engt} & 
\multicolumn{1}{|p{3.98cm}}{1.9493} & 
\multicolumn{1}{|p{3.98cm}}{0.5178} & 
\multicolumn{1}{|p{3.98cm}|}{0.000187} \\ 
\hline
\end{tabular}
\end{adjustbox}
\caption{Multiple regression model to predict module grade}
\end{table}

Following the suggestion of Gelman $\&$ Park (2009), we compare the upper third (or tertile) with the lower tertile rather than a median split. (This avoids the problem of adjacent values being divided by the median boundary.) We summarise this in Table 9. This illustrates dramatically different outcomes between the upper and lower tertiles. This can be expressed quantitatively as an odds ratio of $\sim$20 with 95$\%$ CI [2.6, 152.4] which, despite the broad CI due to failure being a rare event, provides compelling evidence of substantially better odds of passing for students in the top tertile compared with the lowest.

\begin{table}[!htbp]
\renewcommand{\arraystretch}{1.3}
\begin{adjustbox}{max width=\textwidth}
\begin{tabular}{p{5.33cm}p{5.3cm}p{5.28cm}}
\hline
\multicolumn{1}{|p{5.43cm}}{W1-3 Online Engagement Tertile} & 
\multicolumn{1}{|p{5.43cm}}{Failure } & 
\multicolumn{1}{|p{5.43cm}|}{Pass} \\ 
\hline
\multicolumn{1}{|p{5.43cm}}{Lower} & 
\multicolumn{1}{|p{5.43cm}}{18} & 
\multicolumn{1}{|p{5.43cm}|}{146} \\ 
\hline
\multicolumn{1}{|p{5.43cm}}{Middle} & 
\multicolumn{1}{|p{5.43cm}}{6} & 
\multicolumn{1}{|p{5.43cm}|}{158} \\ 
\hline
\multicolumn{1}{|p{5.43cm}}{Upper} & 
\multicolumn{1}{|p{5.43cm}}{1} & 
\multicolumn{1}{|p{5.43cm}|}{163} \\ 
\hline
\end{tabular}
\end{adjustbox}
\caption{Contrasting Failure Rates between W1-3 Online Engagement Terciles for all cohorts}
\label{tab:contrasting_failure_rates_w13_online}\end{table}
In contrast, we also briefly considered the potential for a machine learning classifier (based on rule induction) to discriminate between passing and failing students. The main obstacle for a learning algorithm is the highly imbalanced training data. Recall there are only $\sim$5$\%$ (25/492) instances of the ‘positive’\footnote{ In common parlance failing would not be seen as positive, however, it is usual in the machine learning community to refer to the class of interest as the positive class.} case which creates major challenges for a learning algorithm \cite{he2009learning}. Consequently, machine learning algorithms overfit to the dominant class. Indeed, a naïve rule (i.e., all students pass) would achieve 95$\%$ accuracy. However, this is obviously not a helpful model. Consequently, it was decided not to pursue learning algorithms for pragmatic reasons since rare instances of failure (the positive cases) exacerbated by standard hold-out strategies such as m x n-fold cross-validation mean we would need to deploy over-sampling techniques or some other means to create synthetic training data.

 So, to summarise, yes we can predict – but to a limited extent. Failure and the best predictor variables are based on engagement, particularly those relating to Weeks 1-3 of the module. Essentially, we observe a relationship within the sample; however, as the basis for making an accurate prediction for the individual student, there is a considerable chance of making an erroneous prediction.

\textbf{RQ2: Is engagement linked to final achievement? If so, can we predict students’ final grades from the way they approached the course? }

Next, we broaden out this analysis to consider module outcome in terms of final grade (which is a 17-point scale) as opposed to the binary outcome of RQ1. To do this we build a predictive model based on multiple OLS regression to estimate the OverallGrade. Unfortunately, even our best resulting multiple regression model does not have strong explanatory value since the Adjusted R-squared is only 16$\%$. However, we do note that Weeks 1-3 online engagement remains the most influential part of the model (other than the intercept). 

\begin{figure}[!htbp]
\includegraphics[width=9.61cm,height=6.66cm]{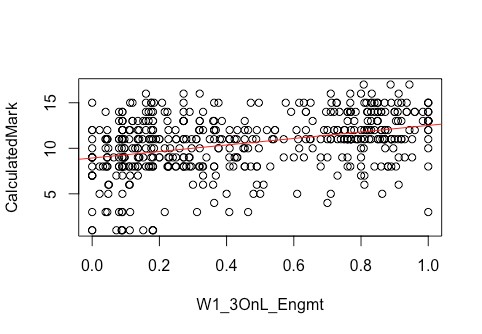}
\caption{Scatterplot of Weeks 1-3 online engagement and CalculatedMark, with the regression line shown in red RQ3 : How does failing the coursework first time impact the final outcome?}
\label{fig:scatterplot_weeks_13_online_engagement}
\end{figure}

Focusing on Weeks 1-3, online engagement can be visualised as a scatterplot with the regression line added in red (see Figure \ref{fig:scatterplot_weeks_13_online_engagement}). This reveals that there is a positive relationship between engagement and grade, but it is weak so we see many examples of students who did not engage yet do well and vice versa. 

\textbf{RQ3: How does failing the coursework first time impact the final outcome? }

To address our third research question, we ask whether those who were reassessed in the coursework could do as well as those who passed first time. We might speculate that they do better because of the extra work required to ‘retrieve’ their coursework, or perhaps they continue to struggle with the module. 

\begin{table}[!htbp]
\renewcommand{\arraystretch}{1.3}
\begin{adjustbox}{max width=\textwidth}
\begin{tabular}{p{3.01cm}p{3.02cm}p{3.2cm}}
\hline
\multicolumn{1}{|p{3.01cm}}{Overall module result } & 
\multicolumn{1}{|p{3.02cm}}{Initial course-work fail } & 
\multicolumn{1}{|p{3.2cm}|}{Initial course-work pass } \\ 
\hline
\multicolumn{1}{|p{3.01cm}}{Fail } & 
\multicolumn{1}{|p{3.02cm}}{25 } & 
\multicolumn{1}{|p{3.2cm}|}{0 } \\ 
\hline
\multicolumn{1}{|p{3.01cm}}{Pass } & 
\multicolumn{1}{|p{3.02cm}}{91 } & 
\multicolumn{1}{|p{3.2cm}|}{376 } \\ 
\hline
\end{tabular}
\end{adjustbox}
\caption{The relationship between initial coursework failure and final module outcome}
\end{table}

Table 10 reveals that $\sim$27$\%$ of students who failed their coursework at the first attempt went onto fail the module outright (i.e., they were unable to pass the coursework on a second attempt). Note that passing the coursework automatically led to an overall module minimum pass (D-). 

\begin{figure}[!htbp]
\centering
\includegraphics[width=5.5cm,height=3.72cm]{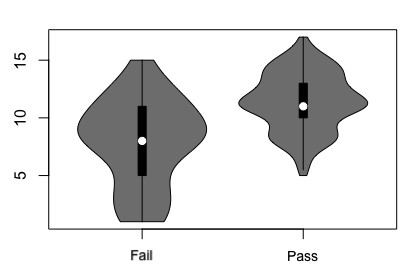}
\caption{Violin plots of the distribution of final grades for students who failed their coursework at the first attempt and those who passed}
\label{fig:violin_plots_distribution_final_grades}
\end{figure}

In terms of final grade, Figure \ref{fig:violin_plots_distribution_final_grades} reveals that some students who failed their coursework at the first attempt were able to redeem themselves and obtained a final grade to a maximum observed grade point of 15 (A). However, clearly the median outcome (denoted by a white circle) is less than that of the students who passed at the first attempt. 

\textbf{RQ4: How stable are patterns over time?}

Our dataset comprises three annual cohorts so we can observe changes between cohorts, such as improvement over time and the impact of other interventions \cite{holte1993very}). We answer this fourth research question in two parts: (i) how do some of the descriptive and summary data change by year (e.g., pass rates); and (ii) how do the predictive models change, in other words how well would a predictive model learned from year \textit{t} work for year \textit{t+1}?

First, consider the pass and failure rates by cohort (see Table 11). It is apparent that the failure rate for 2015/16 is much better/lower and seems quite distinct from 2014/15 or 2016/17.

\begin{table}[!htbp]
\renewcommand{\arraystretch}{1.3}
\begin{adjustbox}{max width=\textwidth}
\begin{tabular}{p{1.55cm}p{1.42cm}p{0.93cm}p{0.93cm}p{1.0cm}p{1.15cm}p{0.99cm}p{1.17cm}}
\hline
\multicolumn{1}{|p{1.77cm}}{} & 
\multicolumn{1}{|p{1.47cm}}{} & 
\multicolumn{2}{|p{1.86cm}}{\textbf{Overall Module }} & 
\multicolumn{2}{|p{1.99cm}}{\textbf{Coursework 1st attempt }} & 
\multicolumn{2}{|p{2.04cm}|}{\textbf{Coursework re-assessment }} \\ 
\hline
\multicolumn{1}{|p{1.77cm}}{\textbf{Year}} & 
\multicolumn{1}{|p{1.47cm}}{\textbf{Cohort size}} & 
\multicolumn{1}{|p{0.99cm}}{Fail} & 
\multicolumn{1}{|p{0.87cm}}{Pass} & 
\multicolumn{1}{|p{0.92cm}}{Fail} & 
\multicolumn{1}{|p{1.07cm}}{Pass} & 
\multicolumn{1}{|p{0.93cm}}{Fail} & 
\multicolumn{1}{|p{1.11cm}|}{Pass} \\ 
\hline
\multicolumn{1}{|p{1.77cm}}{2014-15 } & 
\multicolumn{1}{|p{1.47cm}}{181 } & 
\multicolumn{1}{|p{0.99cm}}{14 } & 
\multicolumn{1}{|p{0.87cm}}{167 } & 
\multicolumn{1}{|p{0.92cm}}{48 } & 
\multicolumn{1}{|p{1.07cm}}{133 } & 
\multicolumn{1}{|p{0.93cm}}{10 } & 
\multicolumn{1}{|p{1.11cm}|}{33 } \\ 
\hline
\multicolumn{1}{|p{1.77cm}}{2015-16 } & 
\multicolumn{1}{|p{1.47cm}}{166 } & 
\multicolumn{1}{|p{0.99cm}}{1 } & 
\multicolumn{1}{|p{0.87cm}}{165 } & 
\multicolumn{1}{|p{0.92cm}}{22 } & 
\multicolumn{1}{|p{1.07cm}}{144 } & 
\multicolumn{1}{|p{0.93cm}}{1 } & 
\multicolumn{1}{|p{1.11cm}|}{20 } \\ 
\hline
\multicolumn{1}{|p{1.77cm}}{2016-17 } & 
\multicolumn{1}{|p{1.47cm}}{145 } & 
\multicolumn{1}{|p{0.99cm}}{10 } & 
\multicolumn{1}{|p{0.87cm}}{135 } & 
\multicolumn{1}{|p{0.92cm}}{46 } & 
\multicolumn{1}{|p{1.07cm}}{99 } & 
\multicolumn{1}{|p{0.93cm}}{10 } & 
\multicolumn{1}{|p{1.11cm}|}{35 } \\ 
\hline
\multicolumn{1}{|p{1.77cm}}{\textbf{Sum }} & 
\multicolumn{1}{|p{1.47cm}}{492 } & 
\multicolumn{1}{|p{0.99cm}}{25 } & 
\multicolumn{1}{|p{0.87cm}}{467 } & 
\multicolumn{1}{|p{0.92cm}}{116 } & 
\multicolumn{1}{|p{1.07cm}}{376 } & 
\multicolumn{1}{|p{0.93cm}}{21} & 
\multicolumn{1}{|p{1.11cm}|}{78 } \\ 
\hline
\end{tabular}
\end{adjustbox}
\caption{Cohort pass and failure rates by year}
\label{tab:table_it_is_apparent_that}\end{table}

\begin{figure}[!htbp]
\includegraphics[width=12.0cm,height=9.81cm]{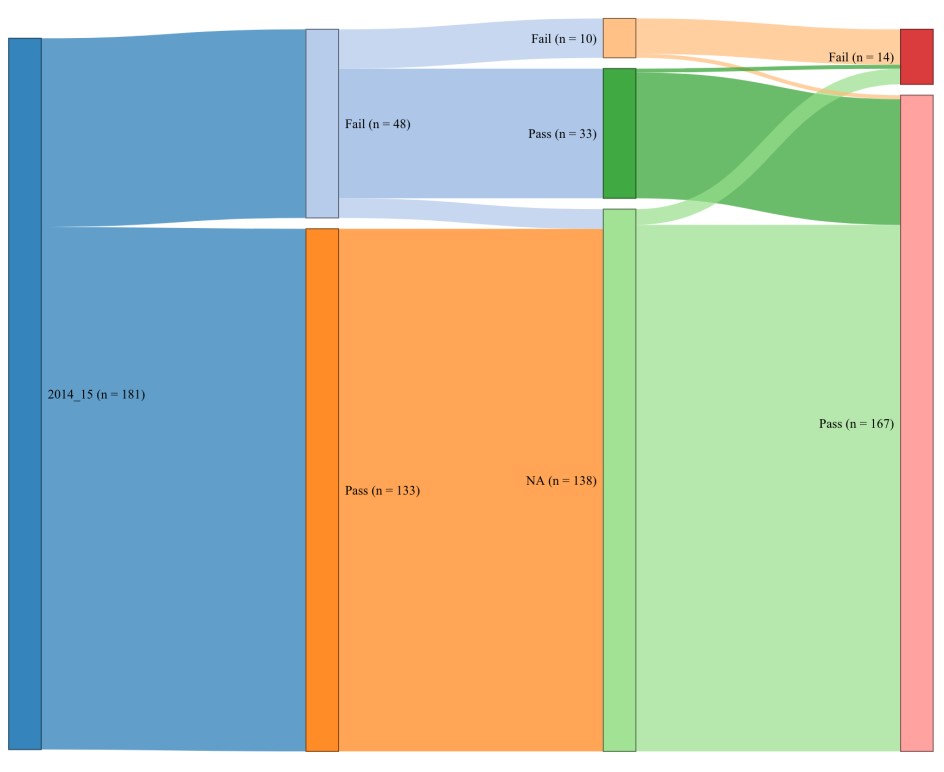}
\caption{Sankey Diagram for 2014/15 cohort}
\label{fig:sankey_diagram_201415_cohort}
\end{figure}

\begin{figure}[!htbp]
\includegraphics[width=12.0cm,height=9.61cm]{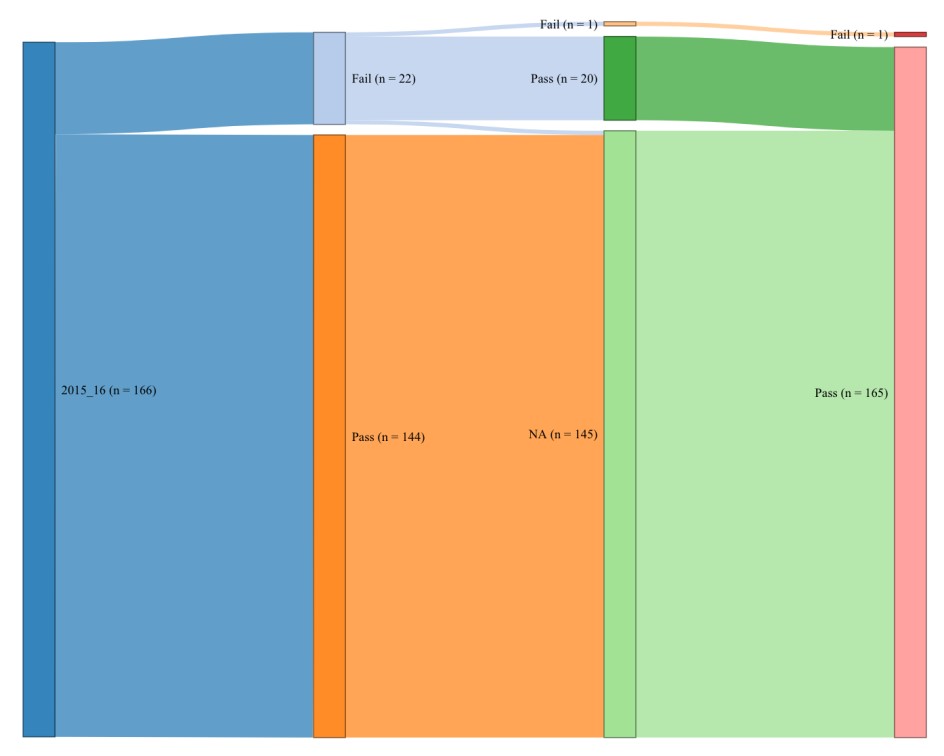}
\caption{Sankey Diagram for 2015/16 cohort}
\label{fig:sankey_diagram_201516_cohort}
\end{figure}

\begin{figure}[!htbp]
\includegraphics[width=12.0cm,height=9.55cm]{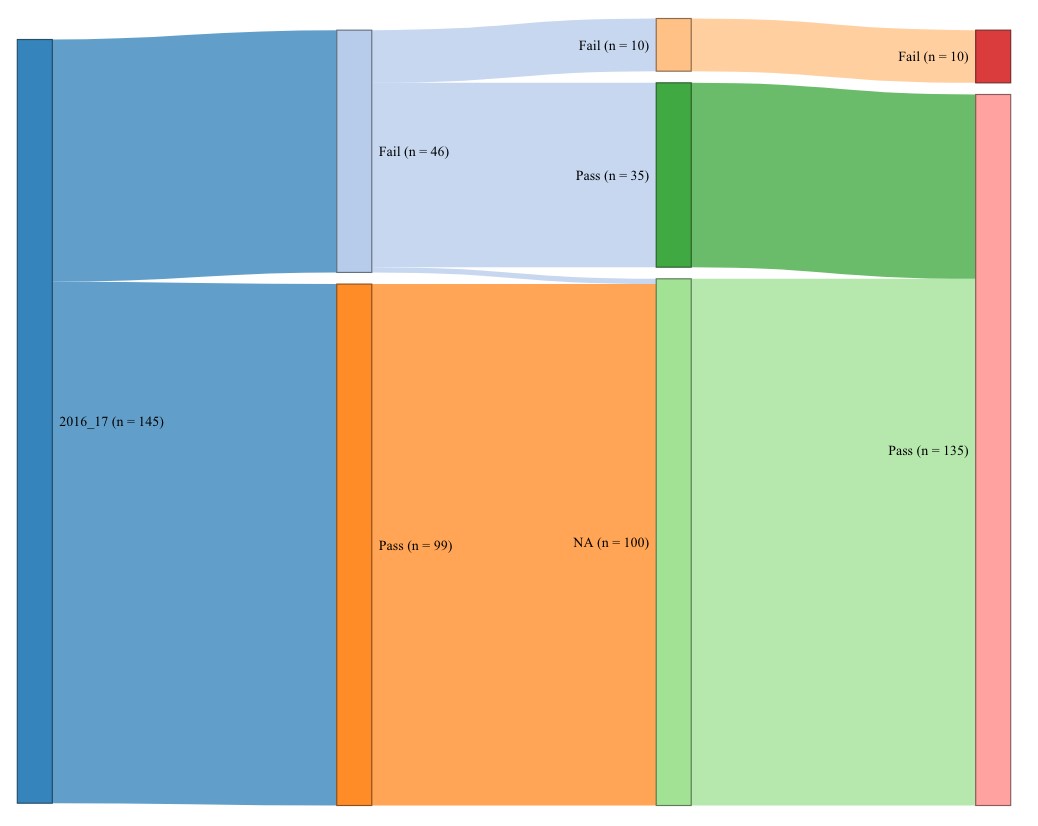}
\caption{Sankey Diagram for 2016/17}
\label{fig:sankey_diagram_201617_cohort}
\end{figure}

Table 11 is easier to visualise as three Sankey Diagrams, one per cohort (see Figures \ref{fig:sankey_diagram_201415_cohort}-\ref{fig:sankey_diagram_201617_cohort}). These diagrams reveal the flows (the widths of which are proportional to the numbers) between stages in the module. So, we see, for instance, that for 2014/15 there 181 students who enrol and of these 133 pass the coursework at the first attempt (and so automatically gain a minimum module pass). Of the remaining 48 students who fail at the first attempt, 33 pass at the second attempt and so also gain a minimum module pass. Ten students fail at the second attempt; however, one of these students had approved extenuating circumstances and so also passed. The remaining students fail, as do five students who did not take up the second re-assessment opportunity. This pattern differs considerably from the 2015/16 cohort.

\begin{figure}[!htbp]
\includegraphics[width=11.57cm,height=10.41cm]{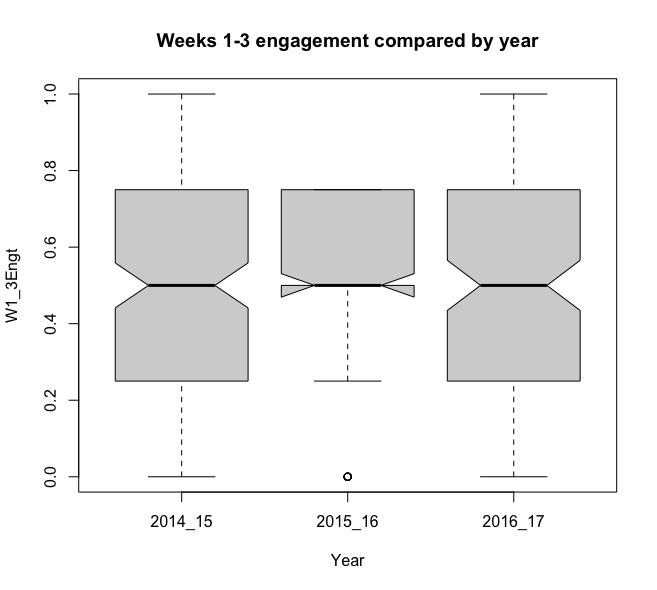}
\caption{Violin plots of cohort grade distribution for the three cohorts}
\label{fig:violin_plots_cohort_grade_distribution}
\end{figure}

Figure \ref{fig:violin_plots_cohort_grade_distribution} shows the violin plots for the three cohorts.

\begin{figure}[!htbp]
\includegraphics[width=0.45\textwidth]{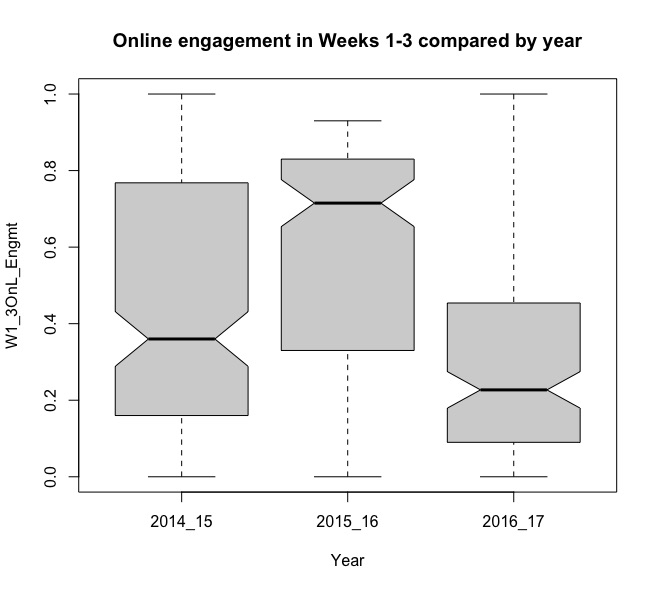}
\includegraphics[width=0.45\textwidth]{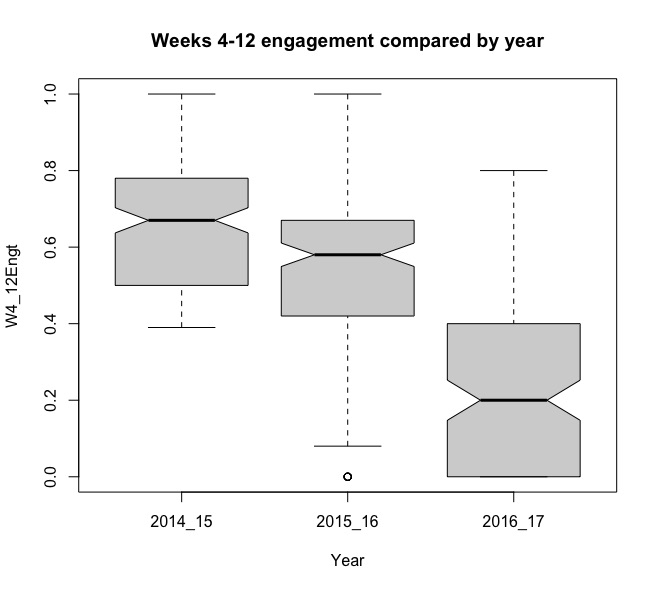}\\
\centering
\includegraphics[width=0.45\textwidth]{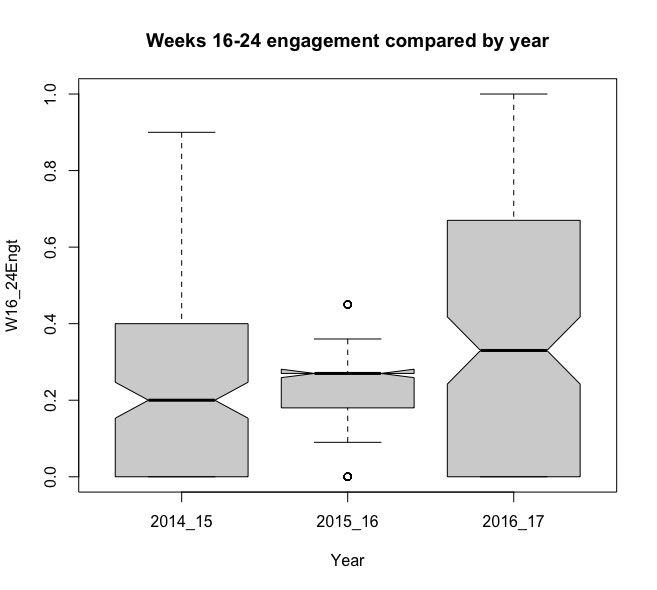}
\caption{Patterns of engagement by year in the different time frames: weeks 1-3; weeks 4-12; weeks 16-24}
\label{fig:violin_plots_engagement}
\end{figure}

Looking in more detail at engagement by cohort, quite distinct patterns emerge. As can be seen in Figure \ref{fig:violin_plots_engagement}, it is interesting to note that the patterns of engagement vary considerably between years, even though the core teaching team remained unchanged. In particular 2015/16 seems quite distinctive. 

\begin{table}[!htbp]
\renewcommand{\arraystretch}{1.3}
\begin{adjustbox}{max width=\textwidth}
\begin{tabular}{p{3.34cm}p{1.66cm}p{1.25cm}p{1.75cm}}
\hline
\multicolumn{1}{|p{3.34cm}}{\textbf{Logistic regression model }} & 
\multicolumn{1}{|p{1.66cm}}{\textbf{2014-15 }} & 
\multicolumn{1}{|p{1.25cm}}{\textbf{2015-16 }} & 
\multicolumn{1}{|p{1.75cm}|}{\textbf{2016-17 }} \\ 
\hline
\multicolumn{1}{|p{3.34cm}}{(Intercept) } & 
\multicolumn{1}{|p{1.66cm}}{-0.647 } & 
\multicolumn{1}{|p{1.25cm}}{n.a. } & 
\multicolumn{1}{|p{1.75cm}|}{1.638 } \\ 
\hline
\multicolumn{1}{|p{3.34cm}}{W1\_3Engt  } & 
\multicolumn{1}{|p{1.66cm}}{3.480 } & 
\multicolumn{1}{|p{1.25cm}}{} & 
\multicolumn{1}{|p{1.75cm}|}{-0.494 } \\ 
\hline
\multicolumn{1}{|p{3.34cm}}{W1\_3OnL\_Engmt } & 
\multicolumn{1}{|p{1.66cm}}{0.656 } & 
\multicolumn{1}{|p{1.25cm}}{} & 
\multicolumn{1}{|p{1.75cm}|}{ 3.215 } \\ 
\hline
\multicolumn{1}{|p{3.34cm}}{W4\_12Engt } & 
\multicolumn{1}{|p{1.66cm}}{3.869 } & 
\multicolumn{1}{|p{1.25cm}}{} & 
\multicolumn{1}{|p{1.75cm}|}{0.327 } \\ 
\hline
\multicolumn{1}{|p{3.34cm}}{W16\_24Engt  } & 
\multicolumn{1}{|p{1.66cm}}{3.385 } & 
\multicolumn{1}{|p{1.25cm}}{} & 
\multicolumn{1}{|p{1.75cm}|}{1.521 } \\ 
\hline
\end{tabular}
\end{adjustbox}
\caption{Comparison of logistic regression models by student cohort year. Note there was only one failure for 2015/16 so the algorithm does not converge}
\label{tab:comparison_logistic_regression_models_student}\end{table}
Next, consider how predictive models (we examine logistic regression models for overall pass or failure) vary by year. If analysis is undertaken on a year-by-year basis, as would normally be the case, then models seem to lack stability (see Table 12). Substantial changes can be seen in the relative importance of explanatory variables and it is not obvious that a model from one cohort would be stationary (stable) over successive years. An additional minor observation is the difficulty of modelling rare events (i.e., failure). This reveals a challenge in building models from one cohort to predict for a subsequent cohort, and the dangers of over-fitting. However, this finding may not be entirely surprising when one considers the many external\footnote{ External, that is external to the predictive model. } sources of variation to student performance in a university setting. 

\section{Discussion of Research Questions 1-4: Issues Identified and Suggestions for Module Leaders}

Having presented answers to the four research questions based on analysis of the collected data, this section will, in line with the overall aim of the paper, discuss findings from the study and provide suggestions for practical steps that module leaders might take to improve student engagement, retention and outcomes. Insights will also be presented into how such bottom-up activities might inform institutional, or top-down, planning in the use of relevant technologies. 

Drawing on the findings in relation to Research Question 1, the approach taken in the module considered in this study highlights students in danger of failing and provides indicators against which to intervene. The clearest indicators are around the early weeks of the module, which showed the strongest correlation between early engagement and final grade (see Figure \ref{fig:correlogram_student_engagement_factors_and}) and where those who failed to engage with the VLE within three weeks were much more likely to fail (see Table 7). 

This leads us to recommend that module leaders should strongly promote the importance of early engagement, explaining its value to their students, and build engagement monitoring and follow-up into the early stages of their modules. 

Considering the findings in relation to Research Question 2, regression analysis shows a positive, if weak, relationship between engagement and final grade. Since overall engagement is not a strong predictor of final grade, though, the key point to stress is that early engagement seems to be the most influential feature in final grade outcome. The ability to analyse different types of engagement in terms of their relationship to grade outcome is, we feel, useful and putting in place systems to gather different types of engagement data is something that we would recommend to module leaders. 

The analysis in relation to Research Question 3 showed that over a quarter of the students who failed the coursework first time went on to fail again (see Table 9), and that the final grade profile of those who failed the coursework first time showed lower overall performance than those who passed first time, though some students who failed first time still achieved A-grades in the module. Beyond this, without more evidence, we do not feel that we can make any concrete recommendations arising from our analysis in relation to this research question.

The stability of patterns over time (Research Question 4) is more nuanced. The median grade in all three years was remarkably constant (see Figures 1 and 2), despite modest changes to the module introduced through the Departmental review process. However, the sub-distributions varied widely from year to year (see Figures 1 and 2), and may account for the anomalously low failure rate of the 2015/16 cohort (see Figures 7-9). Anecdotally, some team members thought the excellent outcomes in 2015/16 represented the sum of all the improvements made up to that date and that the method had been ‘cracked,’ but the 2016/17 outcomes suggested otherwise. 

A consideration to take away from the findings in relation to RQ4 is that there is evidence around identifiable sub-cohorts within the overall cohort and that such identification may encourage module leaders to consider ways to support the sub-cohorts in different ways. 

Stepping back from the specific research questions, the main points to take from the data collection and analysis is that it is possible for module leaders to collect diverse types of information and to integrate the data effectively. Using data collection and analysis in this way makes very clear the presence of sub-groups within a cohort, and how they are progressing and performing. It also provides analytical tools which can help to identify criteria for intervention. 

We have shown how existing methods may be applied, even where low rates of failure, for instance, introduce unhelpful asymmetries. We feel that the most significant message for module leaders, however, is to define as clearly as possible before the start of a run of a module which data fit with the strategy for running the module, and then collect them consistently from the outset. Our experiences show a significant need for, and reliance on, data cleansing which could have been at least reduced had we better understood the importance of particular data items. 

A related point is the importance of institutional systems for gathering data and the ease with which data may be extracted and integrated with other datasets to give a fuller picture of a student’s engagement. Our experiences of finding poor quality and/or inconsistent data in institutional systems may or may not be common, but we see little evidence of institutions paying attention to the nuances of student engagement and outcome tracking associated with individual programmes, modules and cohorts. The result may be centrally-provided systems that have limited value at programme and module level and around, rather than through, which programme and module leaders have to work in order to effectively monitor student progress. 

This may be an artefact of the ‘top-down versus bottom-up’ clash identified in the literature, with institutional systems concentrating on purposes other than direct management of the study cohort at a module level. For instance, the relatively poor quality student records data that we encountered in the first weeks of Term 1 each year have little importance at that point in the academic cycle in terms of key institutional audit and student fee billing processes. What matters institutionally is that the data are correct by the national census date, but the data quality was critical for our purposes at module level because we were using it to identify and follow-up those who had not engaged.

\section{Discussion of Research Questions 5 and 6: Replication and Generalisability}

The wider research questions were:

RQ5: To what extent could this trial be replicated elsewhere?

RQ6: How generalisable are the findings likely to have been?

Though the specific research questions and the technology used to capture attendance online or in person could have been combined in any number of ways to yield module-specific findings, we believe that the approach that was taken can be replicated, at least to some extent. One of the findings of this research is that it is possible to collect good data even when the available systems have not been set up to support the line of enquiry chosen. Against the criticism that technology is in danger of overriding the pedagogy, we have shown how the data collected can be made to fit the requirements of the pedagogy, rather than the other way around. To this extent, we see no reason why others cannot replicate this broad approach. 

In terms of detailed replication, the fact that the module was in a Computer Science department clearly made it easier for the departmental staff both to teach and analyse the outcomes. However, we have shown that there are a range of levels at which the data may be analysed to practical good, and there is a wide range of methods that have been applied. 

The approach taken to identifying and making changes at the module level, through regular reflections on the ‘run’ of a module, are a normal part of the academic cycle. Replicating the approach taken in this study in other modules would obviously depend on the module sharing features in terms of structure, cohort size, and the like; but the study’s value is not necessarily as something to be replicated in a precise way. Rather, it describes an ‘in the wild’ study and presents examples of the application of approaches and analysis techniques in the type of ‘messy’ learning and teaching context that academics face (i.e., where the data are not clean, there may be inconsistencies in data between different systems, there are pressures on data collection, constraints on the activities being undertaken, and various other factors outside the control of those running the module). 

From this perspective, then, the study suggests that it is possible for teaching staff to take much greater control of researching their own learning and teaching context than, in our experience, is normally evident. In principle, it would be possible to for a university to enable such practice-based research, deploying its analytics capability across all modules, and even creating a menu of methods/tools through which staff could collect data and produce findings tailored to their individual requirements, informed by the context of the modules and cohorts that they are teaching. 

Moving on to the issues of generalisability, the specific findings from the study – for instance, the three week rule – may or may not prove to be generally true, but the generalisability of these findings is limited for two reasons. First, by definition we set out to study the application of analytics to a specific module. Second, this is an exploratory observational study. The analyses have not been tested by running experiments where we manipulate treatments to isolate the causal effects. Both the practicalities and ethics of running experiments, even quasi-experiments, are however, likely to present formidable barriers. Other studies could, though, be set up to explore specific findings in as controlled a way as possible. 

The most suggestive findings to consider from this study, should such experiments be planned, are around early engagement and the best ways to promote early and intense engagement by as many members of the cohort as possible. Simply insisting that every student logs in during the first week, and seeing this as a measure of early engagement, for instance, is unlikely to be sufficient, and may obscure identification of students who are not engaging in a meaningful way. Running studies that separate and explore the relative contribution of face-to-face engagement from on-line engagement would also be interesting, though challenging. 

However, the finding that we would argue may be generalised is based more on the high-level approach taken, which leads us to suggest that it is possible to gather relevant information for every cohort and then to apply it to improve the quality of teaching and learning for staff and students. Were others to undertake similar studies, a subsequent meta-analysis could reveal whether there are invariants around engagement with the VLE, progression and recovery from failure that apply regardless of delivery approach, size of class or subject area. For now, though, we note that it is possible to gather – and analyse – such data using the types of technology to which most in Higher Education have everyday access.

\section{Conclusions}

In conclusion, through the analysis of data from a single module over a three-year period, we have presented evidence that may be useful to other module leaders interested in simple ways in which students who are more likely to fail were identified and how different phases of the module related to the final grade depending on students’ engagement with them. The study also highlights the need for on-going work in this area. There would be a value in controlled experiments to allow the exploration of causality between factors, but in most cases there are significant impracticalities to such research. As such, despite their limitations, we would suggest that studies of the type reported in this paper are valuable to the research and practice communities in Higher Education. 

The approach to data collection and analysis presented in this paper provides one way for module designers/leaders to examine hypotheses and refine their teaching practice with the aim of enhancing student achievement from year to year. The up-front planning of the data collection is, though, critical and, in our experience, the effort to create an integrated dataset may be sizeable owing to a lack of focus in institutional systems on supporting this type of on-going engagement and outcome monitoring at module level. These observations are ones that we believe HEIs should take seriously if they are to harness the potential of module-level research to improve learning and teaching practice and the outcomes of students.

\bibliographystyle{plain}
\bibliography{fulltext}

\end{document}